\documentstyle[12pt]{article}
\def\be{\begin{equation}}
\def\ee{\end{equation}}
\def\ba{\begin{eqnarray}}
\def\ea{\end{eqnarray}}

\def\fun#1#2{\lower3.6pt\vbox{\baselineskip0pt\lineskip.9pt

\ialign{$\mathsurround=0pt#1\hfill##\hfil$\crcr#2\crcr\sim\crcr}}}
\begin{document}

\begin{titlepage}
\null\vspace{-62pt}
\begin{flushright} hep-ph/9511391 \\
UMD-PP-96-48 \\
November 1995
\end{flushright}
\vspace{0.2in}

\centerline{{\large \bf  Simple supersymmetric solution}}

\centerline{{\large \bf  to the strong {\it CP} problem} }

\vspace{0.5in}
\centerline{
Rabindra N. Mohapatra\ \ and \ \ Andrija Ra\v{s}in }
\vspace{0.2in}
\centerline{\it Department of Physics}
\centerline{\it University of Maryland}
\centerline{\it College Park, MD 20742}
\vspace{.7in}
\baselineskip=19pt

\centerline{\bf Abstract}
\begin{quotation}

It is shown that the minimal supersymmetric left-right model can provide 
a natural solution to the strong {\it CP} problem without the need for 
an axion, nor any additional symmetries beyond supersymmetry and parity. 

\vspace{0.4in}


\end{quotation}
\end{titlepage}

\baselineskip=19pt

Quantum chromodynamics, which is extremely succesful in describing strongly 
interacting phenomena both in the low as well as the high energy domain,
has the well-known problem that it can lead to uncontrolled amount of
{\it CP} violation in the flavor conserving hadronic processes. This is
the strong {\it CP} problem \cite{jkim87}. The parameter $\bar{\Theta}$ which
characterizes the strength of these {\it CP-}violating interactions is 
constrained by present upper limits on the electric dipole moment of the
neutron to be less than $10^{-9}-10^{-10}$. Presence of such a small number
in a theory indicates the existence of new symmetries beyond the standard
model of electroweak and strong interactions. Three classes of spontaneously
broken symmetries have, in the past, been advocated as solutions to the strong
CP problem: (i) Peccei-Quinn U(1) symmetry \cite{pecc77}; (ii) 
Parity (or left-right) symmetry of weak interactions \cite{mbeg78}
and (iii) softly broken CP symmetry \cite{geor78}. There also
exist other solutions which use less transparent symmetries to constrain the
form of quark mass matrices into interesting forms thereby suppressing
$\bar{\Theta}$ to the desired level \cite{nels83}. In the absence of 
any experimental evidence for or against any of these solutions, 
one can look for theoretical criteria to reduce the number of 
such possibilities. One criterion discussed in recent years
is to use the lore that unlike local symmetries, all global symmetries
are broken by non-perturbative gravitational effect such as black holes and 
wormholes. Since all our solutions involve new global symmetries, one must
investigate whether in the presence  of these effects, the solution to the 
strong {\it CP} problem remain viable. 
In Ref. \cite{kami92} it was shown that the 
presently invisible axion models \cite{jkim79} are incompatible with the above 
non-perturbative effects essentially due to the fact that the PQ symmetry
breaking scale in this case must be $\approx 10^{10}-10^{12}$ GeV. On
the other hand, it was shown in Ref. \cite{bere93} that as long as the 
scales of {\it P} or {\it CP} violation are less than some intermediate
scale, the non-perturbative Planck scale effects do not destabilize 
the second and third solutions to the $\bar{\Theta}-$problem.
	In this letter, we will show that in a class of minimal supersymmetric 
models recently dicussed \cite{cvet84,kuch93} in order
to have automatic {\it R-}parity conservation prior to symmetry breaking, the
strong {\it CP} parameter $\bar{\Theta}$ naturally vanishes both at the
tree and one-loop level, thus providing a solution to the strong 
{\it CP} problem. No additional symmetries are needed for the purpose. 
The only difference between earlier SUSY left-right models and ours
is the inclusion of dimension four Planck scale induced terms, which
are in general expected to be present\cite{kami92}. This provides a
way to ensure that R-parity remains an exact symmetry in the theory even
after the gauge symmetry is spontaneously broken.
 This in combination with the constraints of parity
invariance on the coupling parameters of the theory lead us to our result
that the model provides a solution to the strong CP problem without
the need for an axion. Since the Yukawa couplings in the model are 
complex, the observed
weak CP-violation in the kaon system is explained via the usual 
Cabibbo-Kobayashi-Maskawa phase in the left-handed 
$W$-coupling (as in the standard
model). Some additional interesting properties of the model are:
(i) including Planck scale effects 
leaves the solution unscathed, as in Ref. \cite{bere93};
(ii) unlike the MSSM and the model of Ref. \cite{kuch93}, R-parity is naturally
conserved to all orders in $ {1 \over M_{Pl}}$, so that the lightest
supersymmetric particle (LSP) remains absolutely stable in this model 
and plays the role of CDM and (iii) the SUSY contributions
to electric dipole moment of the neutron are automatically suppressed,
thereby curing the so-called SUSY CP problem.

To see how parity symmetry really suppresses the $\bar{\Theta}$, let
us start by noting that in an electroweak theory, there are two contributions
to $\bar{\Theta}$ at the tree level:
$\bar{\Theta} = \Theta + {\rm Arg} \det (M_u \, M_d)$,
where $\Theta$ is the coefficient of the $G \, \tilde{G}$ term
in the QCD Lagrangian induced by instanton effects and the second term 
is self-explanatory with $M_u$ and $M_d$ denoting
the up and down quark mass matrices. Since $G \, \tilde{G}$ is odd under 
parity, if the theory is required to be parity invariant, we must have
$\Theta = 0$. The vanishing of the second term is however more tricky. 
In the nonsupersymmetric left-right models based on the gauge group
SU$(2)_L \, \times$ SU$(2)_R \, \times$ U$(1)_{B-L}$\cite{pati74},
the quark masses arise from the following gauge invariant Lagrangian:

\be
{\cal L}_Y = {\bf h}^i_{ab} \bar{Q}_{L,a} \Phi_i Q_{R,b} + h.c. \, ,
\ee
where $Q_a = (u_a,d_a)$ ($a = 1,2,3$ for three generations)
and $\Phi_i$ are bidoublets (2,2,0). In the minimal non-supersymmetric model,
one usually considers one $\Phi$ so that there exists another
bidoublet $\tilde{\Phi} \equiv \tau_2 \Phi^* \tau_2$ leading to
two yukawa matrices ${\bf h}^{(1)}$ and ${\bf h}^{(2)}$. Under left-right
(P) symmetry, one assumes that
$Q_{L,a}  \leftrightarrow Q_{R,a}$ and
$\Phi_i \leftrightarrow  \Phi_i^\dagger $.
It is then easy to show that parity invariance demands that
${\bf h}^{(i)} = { {\bf h}^{(i)} }^\dagger $.
Now, if the ground state had the property that  $<\Phi_i>$ is real
(i.e. the ground state is {\it CP-}conserving) then one would have
hermitean mass matrices implying that the second term in 
$\overline{\Theta}$ above is zero. One would then have obtained 
$\bar{\Theta}_{\rm tree} = 0$.
Unfortunately, without extra symmetries, the most general Higgs
potential in non-supersymmetric left-right model has complex couplings and
therefore the vacuum state is necessarily {\it CP-}violating. 
As an example consider the Higgs system $\Phi$,
$(\Delta_L,\Delta_R)$ \cite{moha81}, where $\Delta_L$ and $\Delta_R$ are left
and right SU(2) triplets respectively with $B \, -\, L \, =\, 2$. 
In this model, all but one scalar coupling in the Higgs potential 
are real but the complex one corresponds to 
$|\lambda| \det \, \Phi ( e^{i\alpha} \Delta_L^\dagger \Delta_L
+ e^{-i\alpha} \Delta_R^\dagger \Delta_R) \, + h.c. $
which induces a complex vacuum expectation value (VEV).
Note now that in the presence of complex VEVs $<\Phi>$, the mass matrix is not
hermitean and at the tree level $\bar{\Theta} \neq 0$ despite the
theory being parity invariant. One therefore  needs new symmetries that 
forbid the above term \cite{mbeg78}. 

\vspace{4mm} 
{\bf The supersymmetric model}
\vspace{3mm}

As already mentioned, the gauge group of the theory is 
SU$(2)_L \, \times$ SU$(2)_R \, \times$ U$(1)_{B-L}$ with quarks and 
leptons transforming as doublets under SU$(2)_{L,R}$ depending
on their chirality as follows:Q (2,1,$+ {1 \over 3}$);
$Q^c$(1,2,$- {1 \over 3}$) ; L (2,1,$- 1$); $L^c$ (1,2,+ 1). The
Higgs fields and their transformation properties are:
$\Phi_{1,2}$ (2,2,0); $\Delta$ (3, 1, +2);
$\bar{\Delta}$(3,1,$- 2$); $\Delta^c$(1,3,- 2); $\bar{\Delta}^c$(1,3,$+ 2$).
The superpotential for this theory is given by (we have suppressed
the generation index):

\ba
W & = & 
{\bf h}^{(i)}_q Q^T \tau_2 \Phi_i \tau_2 Q^c +
{\bf h}^{(i)}_l L^T \tau_2 \Phi_i \tau_2 L^c 
\nonumber\\
  & +  & i ( {\bf f} L^T \tau_2 \Delta L + {\bf f}_c 
{L^c}^T \tau_2 \Delta^c L^c) 
\nonumber\\
  & +  & \mu_{\Delta} {\rm Tr} ( \Delta \bar{\Delta} ) + 
\mu_{\Delta^c} {\rm Tr} ( \Delta^c \bar{\Delta}^c ) +
\mu_{ij} {\rm Tr} ( \tau_2 \Phi^T_i \tau_2 \Phi_j ) \nonumber\\
& + & W_{NR}\, .
\ea  
where $W_{NR}$ denotes non-renormalizable terms arising from Planck
scale physics. Typically, $ W_{NR}=(\lambda/M)[Tr(\Delta^c\tau_m
\overline{\Delta}^c)]^2$ + other terms. Being a Planck scale effect,
it can violate parity symmetry and we assume it does.
At this stage all couplings ${\bf h}^{(i)}_{q,l}$, $\mu_{ij}$,
$\mu_{\Delta}$, $\mu_{\Delta^c}$, ${\bf f}$, ${\bf f}_c$ are 
complex with $\mu_{ij}$, ${\bf f}$ and ${\bf f}_c$ being symmetric matrices. 
The terms that break supersymmetry softly to make the theory 
realistic can be written as

\ba
{\cal L}_{\rm soft} & = & \int d^4 \theta \sum_i m^2_i \phi_i^\dagger \phi_i
                      + \int d^2 \theta \, \theta^2 \sum_i A_i W_i 
     + \int d^2 \bar{\theta} \, {\bar{\theta}}^2 \sum_i A_i^* W_i^\dagger
                        \nonumber\\
     & + & \int d^2 \theta \, \theta^2 \sum_p m_{\lambda_p} 
                 {\tilde{W}}_p {\tilde{W}}_p +
           \int d^2 \bar{\theta} \, {\bar{\theta}}^2 \sum_p m_{\lambda_p}^* 
                 {{\tilde{W}}_p}^* {{\tilde{W}}_p}^* \, . 
\label{eq:soft}
\ea

In Eq. \ref{eq:soft},  ${\tilde{W}}_p$ denotes the gauge-covariant
chiral superfield that contains the $F_{\mu\nu}$-type terms with
the subscript going over the gauge groups of the theory including
SU$(3)_c$. $W_i$ denotes the various terms in the superpotential, 
with all superfields replaced by their scalar components and
with coupling matrices which are not identical to those in $W$.
Eq. \ref{eq:soft} gives the most general set of soft breaking terms
for this model.

To see how $\overline{\Theta}=0$ in the model, let us
choose the following definition of left-right 
transformations on the fields and the supersymmetric variable $\theta$:
$Q \leftrightarrow  {Q^c}^\dagger$; 
$L \leftrightarrow  {L^c}^\dagger$ ;
$\Phi_i  \leftrightarrow  {\Phi_i}^\dagger$;
$\Delta \leftrightarrow {\Delta^c}^\dagger$;
$\bar{\Delta} \leftrightarrow {\bar{\Delta}}^{c\dagger}$;
$\theta \leftrightarrow \bar{\theta}$;
${\tilde{W}}_{SU(2)_L} \leftrightarrow {\tilde{W}}^*_{SU(2)_R}$;
${\tilde{W}}_{B-L,SU(3)_C}  \leftrightarrow  {\tilde{W}}^*_{B-L,SU(3)_C}$.
With this definition of L-R symmetry, it is easy to check that
${\bf h}^{(i)}_{q,l} ={{\bf h}^{(i)}_{q,l}}^\dagger$ ;
$\mu_{ij} = \mu^*_{ij}$;
$\mu_\Delta  = \mu^*_{\Delta}$;
${\bf f} = {\bf f}_c^*$;
$m_{\lambda_{SU(2)_L}}  = m_{\lambda_{SU(2)_R}}^*$;
$m_{\lambda_{B-L,SU(3)_C}}  =  m_{\lambda_{B-L,SU(3)_C}}^*$.
From these constraints,
we see that Yukawa couplings still remain complex whereas all couplings
 involving only bidoublet Higgs fields are real. This is the first step in our
proof that $\overline{\Theta}=0$. 

Now we are ready to look for minima of the Higgs potential to see 
whether $<\Phi_i>$ have phases or not. In discussing this, we must 
first recall the relevant result of Ref. \cite{kuch93} which showed
that in order for the 
ground state to respect electromagnetic gauge invariance, one must
break R-parity, {\it i.e.} $< {\tilde{\nu}}^c > \neq 0$ for at least
one generation. This is not desirable for our purpose since
the $<{\tilde{\nu}}^c>$ VEV will always induce
the VEV of $<{\tilde{\nu}}>$ via the leptonic Yukawa 
couplings. Because of these sneutrino VEVs the 
minimum equations generate a small phase in the bidoublet VEVs, which
will upset the hermiticity of the quark mass matrices leading to non-zero
$\overline{\Theta}$. Thus in order to solve the 
strong {\it CP} problem we need to work with the minimum where
$< {\tilde{\nu}}^c > = 0$. So how does one evade the theorem of Ref.
\cite{kuch93}? Let us recall that the result of Ref. \cite{kuch93} is
valid for the most general renormalizable superpotential of the model.
However, if one assumes that non-perturbative Planck scale effects
can induce operators with dimension 4 or higher, the result of 
Ref. \cite{kuch93} is easily avoided leading to the charge conserving
minimum with $< {\tilde{\nu}^c} >= 0$. The simplest operator that is helpful is
$ {\lambda \over M_{Pl}} [ {\rm Tr} ( \Delta^c \tau_m \bar{\Delta}^c) ]^2 $.
The main point is that in the absence of the dimension four terms in the 
superpotential, the global minimum of the theory not only conserves
parity but also violates electric charge conservation as soon as it
breaks the gauge symmetry (i.e. has $<\Delta^c >\neq 0$) and is given by
$< \Delta >=< \Delta^c > = {{1}\over{\sqrt{2}}}v\tau_1$ and similarly for
$< \overline{\Delta}>=< \overline{\Delta}^c >= {{1}\over{\sqrt{2}}}v'\tau_1$.
This happens because the D-term vanishes for this charge violating 
minimum\cite{kuch93} whereas it is non-zero for the charge conserving one
for which $< \Delta^c>= v(\tau_1-i\tau_2)/2$ and $< \overline{\Delta}^c >=
v'(\tau_1+i\tau_2)/2$. As soon as the Planck scale
terms are included, it lifts the charge violating minimum higher than
the charge conserving one for a large range of parameters. In typical
singlet hidden sector Polonyi type models, we estimate $v^2-v^{\prime 2}
\approx {{f^2 M^2_{SUSY}}\over{16\pi^2}}$ so that the charge conserving
minimum occurs for $f\leq 4\pi \left( {{4\lambda \mu_{\Delta}v^4}
\over{M_{Pl}M^4_{SUSY}}}\right)^{{1}\over{4}}$.
Here $f$ is one of the leptonic Yukawa couplings defined in Eq.2.
For $\lambda\approx 1$, $\mu_{\Delta}\approx v\approx M_{SUSY}\approx 1 TeV$,
 we get $f\leq 10^{-3}$.
The parity asymmetric nature of this operator is also crucial for obtaining
a parity violating minimum. We also note that $\lambda$ can be chosen
complex and yet the phase it induces in the vevs being of order 
$v^2/ M^2_{Pl}$ are negligible. 

Having chosen the vev with $< \tilde{\nu}^c >=0$, let us now see
whether the vevs of the $\Phi$ field are real as is needed to solve
the strong CP problem. We have carried out a detailed
analysis of the Higgs potential and find that, at the minimum of the 
potential, it is indeed true. It is clear that that the two-bidoublet
SUSY left-right model being discussed
is a special case of the four Higgs extension of the minimal
supersymmetric standard model as far as the doublet Higgs sector is concerned.
The question of spontaneous CP violation in the latter case
 has been recently studied in Ref.\cite{masi95}, where it is shown
that if a general supersymmetric model 
with two pairs of Higgs doublets has no complex parameters in the
doublet Higgs sector, it cannot break {\it CP} spontaneously
for any range of values of the parameters of the Higgs potential.
Since our model in the Higgs sector is a special case of this, it follows
that the vevs of $\Phi_{i}$ must be real. This then implies that the
quark mass matrices are hermitean and therefore $\bar{\Theta} = 0$ 
naturally at the tree level in our model.

Let us now turn to the one loop contribution to the quark mass
matrices to see if they make any contributions to $\bar{\Theta}$.
Because if the quark mass matrices lose their hermiticity 
at the one loop they will induce too large a value for $\bar{\Theta}$.
There are both Higgs and gaugino mediated diagrams
(Figs. \ref{fig:higgs} and \ref{fig:gaugino} 
respectively). The higgs mediated graph contributes
as follows

\be
\delta M_q^H = [ A_{ij} {\bf h}^{(i)} M^{(0)}_q {\bf h}^{(j)} ] \, .
\ee

Here $M_q^{(0)}$ denotes the tree level contribution. Due to the symmetry
property $\mu_{12}=\mu_{21}$ and reality of $\mu_{ij}$, it follows that 
$\delta M_q^H$ is hermitean. As far as the gauge mediated contribution 
is concerned, $\delta M_q^G \propto M_q^{(0)}$. 
Turning to gaugino contributions, since $m_\lambda$ for the 
SU($2)_{L,R}$ can be complex, a careful analysis is needed to see 
what their contribution to $\bar{\Theta}$ is. We find these contributions
come always in pairs for both left and right gauginos, and because of 
the constraint $m_{\lambda_{SU(2)_L}}= m^*_{\lambda_{SU(2)_R}}$ derived
earlier, their complex parts cancel out when the diagrams 
are summed up. Two typical graphs are shown in Figure \ref{fig:gaugino}. 
Therefore the gauge mediated contribution is also automatically hermitean. 
Thus, the total one loop contribution to $\bar{\Theta}$ vanishes.

From the above discussion, we conclude that the lowest 
order contribution to $\overline{\Theta}$ if any can arise only 
at the two loop 
level. Its contribution to $\bar{\Theta}$ can be crudely estimated to be:
\be
\bar{\Theta} \simeq \left( {{m_tm_b} \over V^2_{WK}} \right)
{ 1 \over {(16 \pi^2)^2}} \left( { {\mu_{ij}^2} \over M^2 } \right) I \, .
\ee
For $\mu_{ij} \simeq 10^{-1} M$, this ``primitive" estimate gives
$\bar{\Theta} \simeq 4 \times 10^{-9} I$, where I denotes the value
of the two loop integral. A more careful estimate will 
also bring in small mixing angles, which will further
suppress $\bar{\Theta}$.

An interesting point to note is that 
since in our model the B-L gaugino and gluino mass 
terms are {\it CP}-conserving, the problem of large
neutron electric dipole moment does not exist 
and one has a simple resolution of the SUSY CP problem encountered
in the MSSM.

In summary, we have shown that minimal models that combine supersymmetry and 
parity invariance provide a simple solution to the strong {\it CP} 
problem without the need to invoke any additional symmetries. The
key elements in our proof are: 
(i) the transformation of supersymmetry coordinate 
$\theta \leftrightarrow \bar{\theta}$ under parity and (ii) the
inclusion of the nonperturbative Planck scale suppressed 
operators in the superpotential. The latter ensures that
the ground state of the theory conserves R-parity, which in turn leads to
real vacuum expectation values of the bidoublet fields 
for arbitrary values of the parameters in the theory. This together
with the hermiticity of the Yukawa couplings generic to left-right models
leads to our solution to the strong {\it CP} problem. 

\vskip 0.5cm

\noindent{\bf Note added in proof:} After this work was completed, 
we came across a paper by R. Kuchimanchi \cite{kuch95} which also
arrives at the same result, under the assumption that all gaugino masses
are same at the Planck scale.

\vskip 0.5cm

{\bf Acknowledgments}

\vskip 0.3cm

This work was supported by the NSF grant No. PHY $9421385$.
The work of R.N.M. is also partially suported by the Distinguished
Faculty Research award by the University of Maryland.


\newpage
\textwidth 5.75in
\unitlength=1.00mm
\thicklines
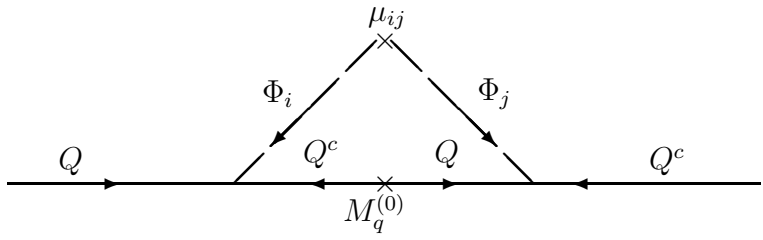
\begin{figure}
\begin{picture}(33.66,182.34)
\put(30.00,120.00){\line(1,0){100.00}}
\multiput(60.00,120.00)(5,5){4}{\line(1,1){4.00}}
\multiput(100.00,120.00)(-5,5){4}{\line(-1,1){4.00}}
\put(40.00,123.00){\makebox(0,0)[rc]{$Q$}}
\put(120.00,123.00){\makebox(0,0)[rc]{$Q^c$}}
\put(74.00,124.00){\makebox(0,0)[rc]{$Q^c$}}
\put(90.00,124.00){\makebox(0,0)[rc]{$Q$}}
\put(83.00,142.00){\makebox(0,0)[rc]{$\mu_{ij}$}}
\put(83.00,116.00){\makebox(0,0)[rc]{$M_q^{(0)}$}}
\put(68.00,132.00){\makebox(0,0)[rc]{$\Phi_i$}}
\put(97.00,132.00){\makebox(0,0)[rc]{$\Phi_j$}}
\put(35.00,120.00){\vector(1,0){10}}
\put(115.00,120.00){\vector(-1,0){10}}
\put(80.00,120.00){\vector(-1,0){10}}
\put(80.00,120.00){\vector(1,0){10}}
\put(81.80,120.00){\makebox(0,0)[rc]{$\times$}}
\put(81.80,139.00){\makebox(0,0)[rc]{$\times$}}
\put(70,130){\vector(-1,-1){5}}
\put(90,130){\vector(1,-1){5}}
\end{picture}
\vspace{-8cm}
\caption{ Higgs contribution to one loop 
calculation of $\bar{\Theta}$.}
\label{fig:higgs}
\end{figure}


\newpage
\textwidth 5.75in
\unitlength=1.00mm
\thicklines
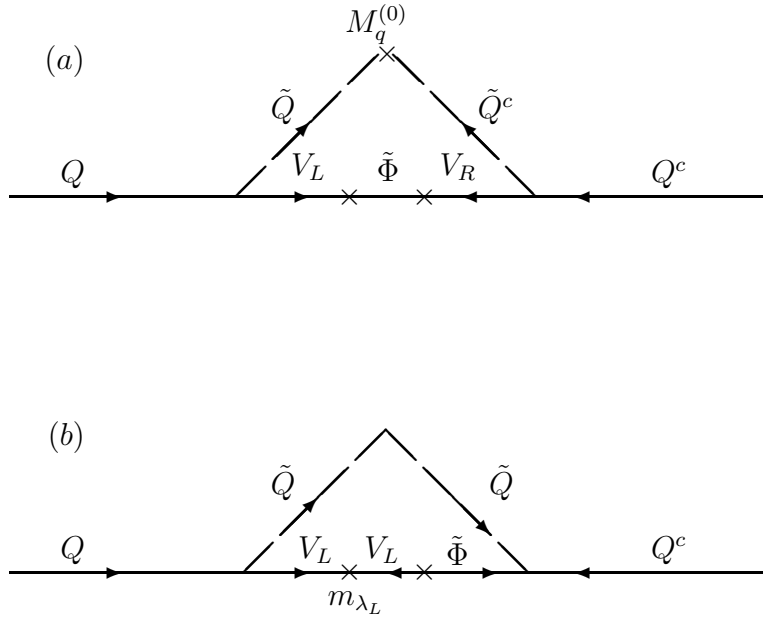
\begin{figure}
\begin{picture}(33.66,182.34)
\put(40.00,188.00){\makebox(0,0)[rc]{$(a)$}}
\put(30.00,170.00){\line(1,0){100.00}}
\multiput(60.00,170.00)(5,5){4}{\line(1,1){4.00}}
\multiput(100.00,170.00)(-5,5){4}{\line(-1,1){4.00}}
\put(83.00,193.00){\makebox(0,0)[rc]{$M_q^{(0)}$}}
\put(40.00,173.00){\makebox(0,0)[rc]{$Q$}}
\put(120.00,173.00){\makebox(0,0)[rc]{$Q^c$}}
\put(72.00,174.00){\makebox(0,0)[rc]{$V_L$}}
\put(81.80,174.00){\makebox(0,0)[rc]{$\tilde{\Phi}$}}
\put(92.00,174.00){\makebox(0,0)[rc]{$V_R$}}
\put(68.00,182.00){\makebox(0,0)[rc]{$\tilde{Q}$}}
\put(97.00,182.00){\makebox(0,0)[rc]{$\tilde{Q}^c$}}
\put(35.00,170.00){\vector(1,0){10}}
\put(115.00,170.00){\vector(-1,0){10}}
\put(60.00,170.00){\vector(1,0){10}}
\put(100.00,170.00){\vector(-1,0){10}}
\put(76.80,170.00){\makebox(0,0)[rc]{$\times$}}
\put(86.80,170.00){\makebox(0,0)[rc]{$\times$}}
\put(81.80,189.00){\makebox(0,0)[rc]{$\times$}}
\put(65,175){\vector(1,1){5}}
\put(95,175){\vector(-1,1){5}}
\put(40.00,138.00){\makebox(0,0)[rc]{$(b)$}}
\put(30.00,120.00){\line(1,0){100.00}}
\multiput(61.00,120.00)(5,5){4}{\line(1,1){4.00}}
\multiput(99.00,120.00)(-5,5){4}{\line(-1,1){4.00}}
\put(40.00,123.00){\makebox(0,0)[rc]{$Q$}}
\put(120.00,123.00){\makebox(0,0)[rc]{$Q^c$}}
\put(73.00,123.00){\makebox(0,0)[rc]{$V_L$}}
\put(91.00,123.00){\makebox(0,0)[rc]{$\tilde{\Phi}$}}
\put(81.80,123.00){\makebox(0,0)[rc]{$V_L$}}
\put(68.00,132.00){\makebox(0,0)[rc]{$\tilde{Q}$}}
\put(97.00,132.00){\makebox(0,0)[rc]{$\tilde{Q}$}}
\put(79.80,116.00){\makebox(0,0)[rc]{$m_{\lambda_L}$}}
\put(35.00,120.00){\vector(1,0){10}}
\put(115.00,120.00){\vector(-1,0){10}}
\put(60.00,120.00){\vector(1,0){10}}
\put(90.00,120.00){\vector(-1,0){10}}
\put(85.00,120.00){\vector(1,0){10}}
\put(76.80,120.00){\makebox(0,0)[rc]{$\times$}}
\put(86.80,120.00){\makebox(0,0)[rc]{$\times$}}
\put(66,125){\vector(1,1){5}}
\put(89,130){\vector(1,-1){5}}
\end{picture}
\vspace{-8cm}
\caption{Examples of gaugino contributions to one loop 
calculation of $\bar{\Theta}$. $V_{L,R}$ are left and right gauginos, 
respectively. The gaugino mass $m_{\lambda_L}$ is in general complex.
There is an analogous graph to b) that involves right-handed gauginos.}
\label{fig:gaugino}
\end{figure}

\end{document}